\documentclass[prl,aps,10pt,twocolumn,showpacs,floatfix,superscriptaddress,amsmath,amssymb,nofootinbib]{revtex4-1}
\usepackage{graphicx}
\usepackage{dcolumn}
\usepackage{bm}
\usepackage{natbib}
\usepackage[normalem]{ulem}
\usepackage{amsmath}
\usepackage{esint}
\usepackage{color}
\usepackage[pdf]{pstricks}

\usepackage{float}
\newcommand{\ba}{\begin{eqnarray}}
\newcommand{\ea}{\end{eqnarray}}
\def\be{\begin{equation}}
\def\ee{\end{equation}}

\begin{document}

\title{Microscopic model for magneto-electric coupling through lattice distortions}

\author{D.C.\ Cabra}
\affiliation{
IFLySiB-CONICET and Departamento de F\'{\i}sica, Universidad Nacional de la Plata,
C.C.\ 67, (1900) La Plata, Argentina}

\author{A.O.\ Dobry}
\affiliation{
Facultad de Ciencias Exactas, Ingenier\'\i a y Agrimensura,
~~\\
Universidad Nacional de Rosario and Instituto de F\'\i sica Rosario,
~~\\
Bv. 27 de Febrero 210 bis, 2000 Rosario, Argentina}

\author{C.J.\ Gazza}
\affiliation{
Facultad de Ciencias Exactas, Ingenier\'\i a y Agrimensura,
~~\\
Universidad Nacional de Rosario and Instituto de F\'\i sica Rosario,
~~\\
Bv. 27 de Febrero 210 bis, 2000 Rosario, Argentina}

\author{G.L.\ Rossini}
\affiliation{
IFLP-CONICET and Departamento de F\'{\i}sica, Universidad Nacional de la Plata,
C.C.\ 67, (1900) La Plata, Argentina
}

\date{\today}

\begin{abstract}
We propose a microscopic magneto-electric model in which the coupling between spins and electric dipoles 
is mediated by lattice distortions. 
The magnetic sector is described by a spin S=1/2 Heisenberg model 
coupled directly to the lattice via a standard spin-Peierls term and 
indirectly to the electric dipole variables via the distortion of the surrounding electronic clouds. 
Electric dipoles are described by Ising variables for simplicity. 
We show that the effective magneto-electric coupling which arises due the interconnecting lattice deformations 
is quite efficient in one-dimensional arrays. 
More precisely, we show using bosonization and extensive DMRG numerical simulations that 
increasing the magnetic field above the spin Peierls gap, 
a massive polarization switch-off occurs due to the proliferation of soliton pairs. 
We also analyze the effect of an external electric field $E$ when the magnetic system is in a 
gapped (plateau) phase and show that the magnetization can be electrically 
switched between clearly distinct values. 
More general quasi-one-dimensional models and two-dimensional systems are also discussed.   
\end{abstract}


\maketitle


Multiferroic materials exhibit a magnetoelectric (ME)
coupling between their electrical and magnetic moments, 
a promising feature for device designs controlling magnetization with electric fields, or conversely
electrical polarization with magnetic fields.
They have been the subject of intense research in the last decade, 
a century later than the pioneering insight of P.\ Curie \cite{Curie-1894} and fifty years after
the first theoretical prediction and experimental realization 
in Cr$_2$O$_3$ \cite{Dzyaloshinskii-1960,Astrov-1961}.
Current revival may be traced back to the discovery of simultaneous polarization and magnetization 
in bismuth ferrite BiFeO$_3$ \cite{BiFeO} and 
gigantic magnetoelectric effects in rare earth perovskite manganites Te(Dy)MnO$_3$ \cite{Kimura-2003}.
Since then a series of exciting new materials and new microscopic descriptions have been developed (see 
the reviews \cite{Fiebig,Khomskii,CheongMostovoy,RameshSpalding-2007,Dong-2015,TokuraR,Khomskii2,Dagotto-2019} and references therein).
Still, technologically useful multiferroic materials are very rare and constitute an active area of research. 

Multiferroics are usually divided into two main groups, named type I and II, 
depending on whether ferroelectricity  and magnetism have different or the same origin (see {\it e.g.} \cite{Khomskii,TokuraR} and references therein). 
Within the second group, in which ferroelectricity occurs in a magnetically ordered state, 
further distinction can be made if the magnetic order is collinear \cite{Khomskii} or non-collinear \cite{Katsura, Dagotto}.     

Previous studies \cite{ANNNI-models,RVO,R2CoMnO6,Dagotto3,Dong-2009,Medarde,Catalano,Chapon} 
have linked magnetostriction effects to magnetoelectricity, 
in particular for quasi-one dimensional materials 
like Ca$_3$CoMnO$_6$ \cite{ANNNI-models}, 
R$_2$V$_2$O$_7$ (R=Ni, Co) \cite{RVO}, 
double perovskites R$_2$CoMnO$_6$ (R=Er, Ho, Tm, Yb, Lu) \cite{R2CoMnO6},
and also for more general cases such as
magnetic E-type  HoMnO$_3$ manganite \cite{Dagotto3,Dong-2009}, 
the nickelate family RNiO$_3$ (R=La, Pr, Nd, Sm, \dots, Lu), see {\em e.g.} \cite{Medarde,Catalano}, 
RMn$_2$O$_5$ manganites (R=Tb-Lu) \cite{Chapon}, 
etc. 

In the present Letter, we focus on quasi-one dimensional materials with collinear magnetic orders 
and propose an effective microscopic model in which the ME coupling is mediated by lattice distortions. 
Our main motivation arises from many different experiments where the coupling between magnetic moments, 
elastic distortions and electric dipoles have been observed, 
in particular, those in \cite{Giovannetti, Streltsov} where multiferroicity has been linked to 
magnetoelastic deformations in collinear spin models, which in turn produce a net electric polarization. 

In this context, we aim to provide a natural microscopic connection between 
the electro-elastic and magneto-elastic effects and the resulting ME coupling. 
To this end we propose a model describing magnetic ions with spin S=1/2, dipolar degrees of freedom and
%
deformations along a preferred axis, 
which allows for a description in terms of almost independent chains of octahedra,
as is the case for {\it e.g.} \cite{Streltsov, ANNNI-models, Dagotto3}, 
or any other structural units.
We find, among other effects, that this model allows for 
a switch-off the electric polarization by applying a magnetic field, 
as well as for a magnetization jump induced by varying an electric field. 
These functionalities are key features that could lead to 
multiferroics based technologies \cite{techno}.

We consider a chain of spin 1/2 magnetic ions \cite{Naka}
with the coupling to the lattice taken for simplicity as an adiabatic spin-Peierls term. 
We also assume that the ions whose motion produce the electric dipoles move in a deep enough double-well potential (the 
so-called order-disorder limit \cite{Bruce})  
so that the orientation of electric dipoles is 
%
%
described by local Ising variables $\sigma_i$. 
Under longitudinal distortions, we assume that dipoles remain located middle-way between magnetic ions. 
This granted, 
the coupling between elastic deformations and electric dipoles recognize two contributions: 
one stems from the natural $1/r^3$ dependence of the dipole-dipole interaction and 
the other, central in our proposal, arises from a pantograph mechanism \cite{Jaime}. 
As changes in longitudinal bond lengths are related to the heights of the basic structural units, 
distortions change the width of the double well potentials which in turn modify the dipolar strengths.
A slight generalization could include the so-called bond-bending effects, 
where the magnetic superexchange is better described in terms of bond angles \cite{Dagotto3}; 
the conclusions of the present work would remain unchanged.

\begin{figure}[ht]
\centering
\includegraphics[width=\columnwidth,keepaspectratio]{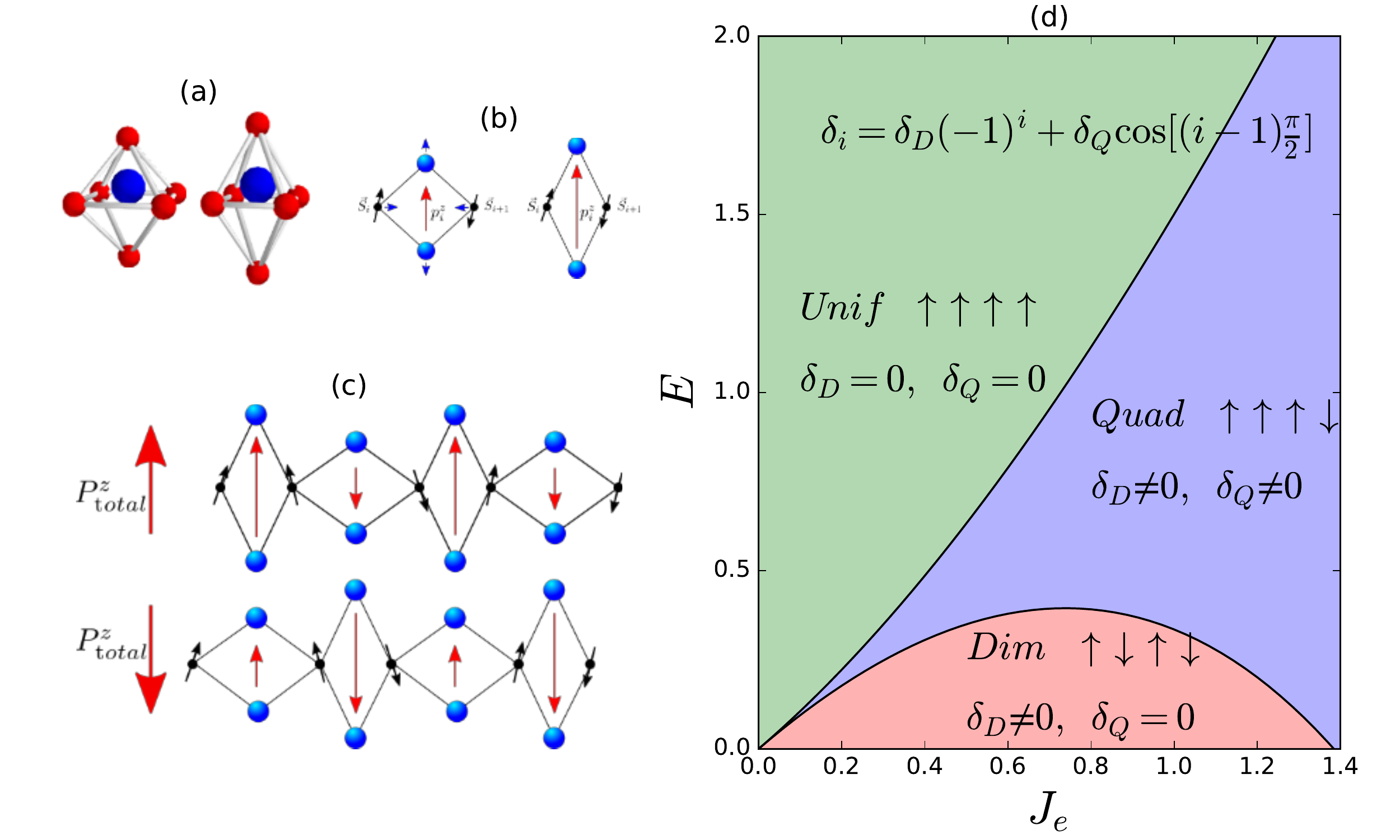} 
 \caption{ Description of the pantograph mechanism and electroelastic phase diagram.
 (a) Typical distortion of basic structural units containing magnetic and electric variables.
 (b) Minimal model linking ion displacements (blue arrows) with spin and dipolar variables (black and red arrows), 
 when a shear strain relates a distance reduction between magnetic ions with a dipolar strength  enlargement. 
(c) The two possible dimerized chain configurations (related by one site translation, $\mathbb{Z}_2$), 
  producing  net polarizations in the presence of an antiferroelectric dipolar phase.
(d) Dipolar phases and electroelastic distortions under an external electric field ($\alpha=0$, $\beta=0.2$). 
}
 \label{fig:1}
\end{figure}

Assuming a preferred direction for the magnetoelastic distortions, 
a minimal geometry for the pantograph mechanism is depicted in Fig.\ \ref{fig:1}(b-c), where for definiteness we set the 
dipoles to be perpendicular to the chain direction; octahedra in three-dimensional materials (a) undergo a similar process.
%
In the following we analyze this simple geometry, considering the Hamiltonian 
\begin{equation}
 H_{ME}=H_{SP}+H_D
 \label{eq:HME}
\end{equation}
where $H_{SP}$ is the usual spin-Peierls Hamiltonian  for S=1/2 spins $\vec{S_i}$ and bond length distortions $\delta_{i}$,
\begin{equation}
 H_{SP}=\sum_i J_m(1-\alpha \delta_i)\vec{S_i}\cdot\vec{S}_{i+1}+\frac{K}{2}\sum_i (\delta_{i})^2
 \label{eq:HSP}
\end{equation}
with antiferromagnetic exchange $J_m$  and elastic stiffness $K$,
and $H_{D}$ is the (long range) electric dipolar energy. 
For transverse uniaxial dipoles $H_{D}$ can be written as
\begin{equation}
 H_D=\lambda_2 \sum_{i<j}\frac{1}{r_{ij}^3}p_i^z p_j^z  \, 
 \label{eq:HD}
\end{equation}
where  the distance
$r_{ij} = r_{ij}(\{\delta_k\})$  between dipoles at links $j>i$ depends on distortions.
The electric dipolar moments also depend on distortions by the pantograph mechanism, leading in a linear approximation to 
\begin{equation}
 p_i^z=p_0(1-\beta \delta_i)\sigma_i \, .
 \label{eq:pz-delta}
\end{equation}
External magnetic and electric fields along the $z$ axis couple to the spins and dipoles, respectively, by
\begin{equation}
H_\text{fields}= -h \sum_i S^z_i -E \sum_i p_i^z \, . 
\label{eq:H-fields}
\end{equation}

%

In a general geometry, both $\alpha$ and $\beta$  should be understood as phenomenological microscopic parameters, that could be fitted by experiments or by first principles computations. The transversality condition on dipole orientations could be relaxed, either because of classical tilting or the inclusion of quantum fluctuations; in these cases our model requires further elaboration, to be reported in a forthcoming work.

%
%
%
In the case that screening makes negligible dipolar interactions beyond first neighbors, the Hamiltonian in Eq.\ (\ref{eq:HME}) simplifies to
\begin{eqnarray}
H_{ME}& = & J_m\sum_i (1-\alpha \delta_i)\vec{S_i}\cdot\vec{S}_{i+1}+\frac{K}{2}\sum_i (\delta_{i})^2 \nonumber \\
&+& J_e \sum_i \left[1-\left(\beta+\frac{3}{2}\right) (\delta_i+\delta_{i+1})\right]\sigma_i\sigma_{i+1} 
\label{eq:H-NN}
\end{eqnarray}
where $J_e = \lambda_2 (p^0)^2$ is the undistorted effective electric exchange coupling. 
Integrating out deformations would lead to a quartic expression
coupling directly the magnetic and elastic degrees of freedom, 
similar to that proposed in \cite{Naka} to describe organic molecular solids. 
The microscopic derivation of this ME coupling will also be the subject of a forthcoming paper. 
We recall that the pantograph effect in Eq.\ (\ref{eq:pz-delta}) and 
the dependence of dipole-dipole electrostatic couplings on distance 
are at the root of the electroelastic coupling mechanism.

The electroelastic part of the Hamiltonian (setting $\alpha=0$) is easily analyzed on classical grounds. 
Distinct dipolar configurations are favoured according to the electric field and the different couplings considered, 
leading to a rich phase diagram. 
We show in Fig.\ \ref{fig:1}(d) the electroelastic phases in the $E-J_e$ plane
for 
$\beta=0.2$; $K$ sets the energy scale. 
The lattice distortions can be analytically computed as a superposition of 
period two and/or period four harmonic distortions.
The dimerized phase ({\em Dim}) has vanishing polarization at $E=0$, 
slightly raising until a critical field $E_{c1}$ where it jumps to nearly half of saturation in a quadrumarized phase ({\em Quad}). 
Distortions have contributions from both harmonics along this phase and the
polarization also raises slightly, until a jump to saturation at a critical field $E_{c2}$.

On the other hand,  the magnetoelastic part of the Hamiltonian (setting $J_e=0$) has been extensively studied 
mainly since the discovery of CuGeO$_3$ \cite{CuGeO} and the spin-Peierls effect is well established: 
the system is unstable to a lattice deformation pattern commensurate with magnetic correlations 
and eventually dimerizes at zero magnetic fields. 
This mechanism happens to be effective also in frustrated chains, 
to give rise to spin gaps (magnetization plateaux) at non-zero magnetization $M$ \cite{Tertions}. 
Magnetic excitations with $S_z=1$ (magnons) on top of a given plateau split into a number of solitons 
which is fixed by the plateau magnetization ratio. 
These solitons repel each other and form hence a periodic array \cite{Lorenz_1998}.
%
An efficient analysis can be made in the bosonization framework
(see \cite{Tertions} for details). 
In this language \cite{Haldane80} the continuum expression for 
the spin energy density $\vec{S_i}\cdot\vec{S}_{i+1} \rightarrow \rho(x)$ reads
\begin{equation}
\rho(x) =a \, \partial_x \phi
 + b :\cos(2k_F x+\sqrt{2\pi}\phi): + \cdots
\label{robos}
\end{equation}
where $\phi$ is the bosonic field, $k_F = \frac{\pi}{2}(1-M)$, $M$ is the magnetization (relative to saturation), $a$, $b$ are
$M$-dependent non-universal constants and the ellipses indicate higher harmonics. 
The magnetoelastic coupling will then be effective when distortion modulations are commensurate with spin energy density oscillations.


Our approach to the full Hamiltonian in Eqs.\ (\ref{eq:H-fields}, \ref{eq:H-NN}) is based on a 
self consistent adiabatic procedure to minimize the energy 
for a given (classical) dipolar and (quantum) spin configuration,  setting distortions as
\begin{eqnarray} 
\delta_i&=&\frac{1}{K} [ J_e(\beta+3/2)<\sigma_i \sigma_{i+1}
+\sigma_{i-1} \sigma_{i}>
\nonumber\\
&+& J_m\alpha <\vec{S_i}\vec{S}_{i+1}>- p_0 E \beta <\sigma_i>]
\label{eq:selfcon}
\end{eqnarray}
(with a subtraction of  their average in order to fulfill a fixed length constraint).
We have performed an iterative numerical analysis based on
the Density Matrix Renormalization Group (DMRG) to solve the magnetic and electric sectors in the adiabatic equations (\ref{eq:selfcon}), along the lines stated in \cite{feiguin1997numerical} and implemented in a similar context in \cite{Tertions}.
%
We have used periodic boundary conditions, keeping $m=300$ states during up to more than 100 sweeps  in the worst cases, getting truncation errors lower than $O(10^{-12})$.


The present model is capable of displaying the ME interplay. 
In particular, when spin-Peierls dimerization occurs at zero magnetic field 
and
the magnetic subsystem is in a gapped phase with $M=0$, 
one has $2k_F=\pi$ and the more relevant modulation term which is commensurate with the spin energy density oscillations reads
%
%
$\delta(x) = \delta_D \cos(\pi  x + q \pi) \ , \ q=0,1$. 


For $E=0$ 
the electric subsystem is in the antiferroelectric Ising regime and exhibits a  spontaneous polarization 
%
$P^z_{total} \equiv \frac{1}{p^0}\sum_i p^z_i = \sum_i \sigma_i(1-\beta \delta_i)  = \pm P_{\text{sp}} $
where $P_{\text{sp}}=\beta \delta_D N $. 
Notice that the polarization is extensive and spontaneous, 
with $\delta_D \neq 0$ 
due to the spin-Peierls effect. 
Moreover, the polarization has two possible orientations depending on the 
breaking of the translational symmetry of the magnetoelastic chain 
into ${\mathbb Z}_2$ as indicated in Fig.\ \ref{fig:1}(c). 
This in turn induces a spontaneous breaking of inversion symmetry along the $z$ axis.
\begin{figure}[hbt!]
\centering
    \includegraphics[width= 1. \columnwidth]{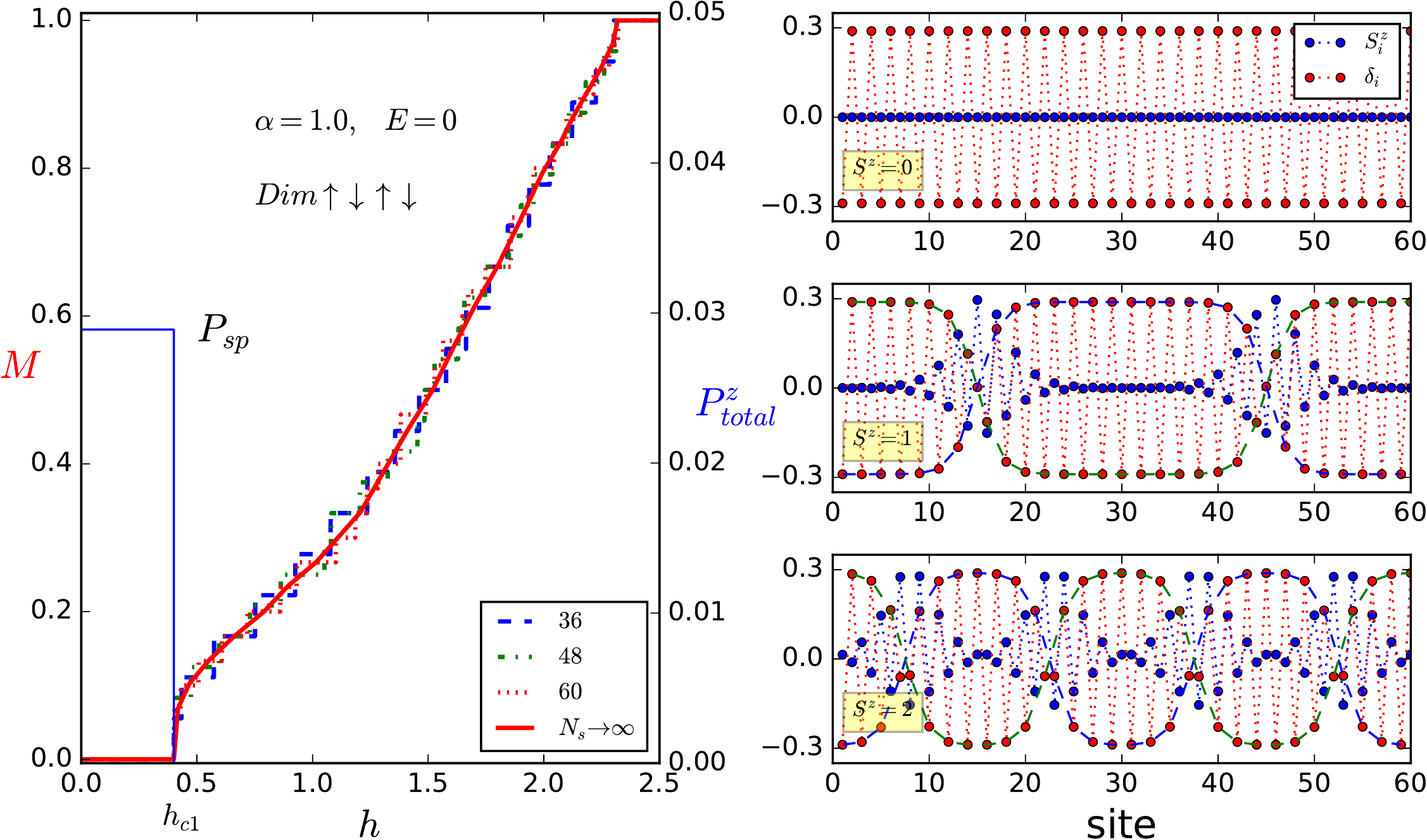}
   \caption{Numerical results for $E=0$ (setting $J_e=0.5$, $\alpha=1$ and $\beta=0.2$).
   Left panel: Magnetization curve and net polarization (relative to saturation).
   A finite magnetic field is necessary to overpass the spin gap, dropping off the spontaneous polarization. 
   Right panels: local magnetization and distortion profiles for $S^z=0$ and the first two magnetized excitations $S^z=1,2$ indicating that equidistant soliton pairs (analytically fitted by dashed lines) proliferate as the magnetic field is increased.  }
\label{fig:2}
\end{figure}
By increasing the magnetic field above the spin gap ($h>h_{c1}$) there occurs  
an incommensurate transition with the excitation of localized singlets  into triplets, which 
decay into pairs of solitons. 
The double degeneracy of dipolar antiferroelectric configurations has a dramatic effect on the polarization: 
as solitons, for $E=0$, form a regular array \cite{Lorenz_1998} interpolating between $q=0,1$, $P^z_{total}$ vanishes identically. 

Thus the magnetic transition causes a complete switch-off of electrical polarization,
%
%
$P^z_{total}(h>h_{c1})  =0$. 
This effect could be observed in inelastic neutron scattering experiments.

 The numerical results shown in Fig.\ \ref{fig:2} illustrate the  polarization switch-off mechanism: 
 the left panel shows the presence of a magnetization plateau at $M=0$ and 
 a critical magnetic field $h_{c1}$ to overcome it; the right panels show the spin and distortion configurations, as well 
as the dipolar background and the net polarization. 
 For $M=0$ the alternating distortions are in phase (say $q=0$) along the chain, while for $S^z=1,2$ well defined equidistant solitons produce regions with $q=0,1$ and a vanishing net polarization; 
 the analytical expression for the first soliton pair,
%
$\delta_i=\pm\delta_D \tanh(\frac{i-i_1}{\xi})\tanh(\frac{i-i_1+N/2}{\xi})$,
%
is indicated with dashed lines in the right middle panel.

\begin{figure}[ht!]
\centering
   \includegraphics[width=1.\columnwidth]{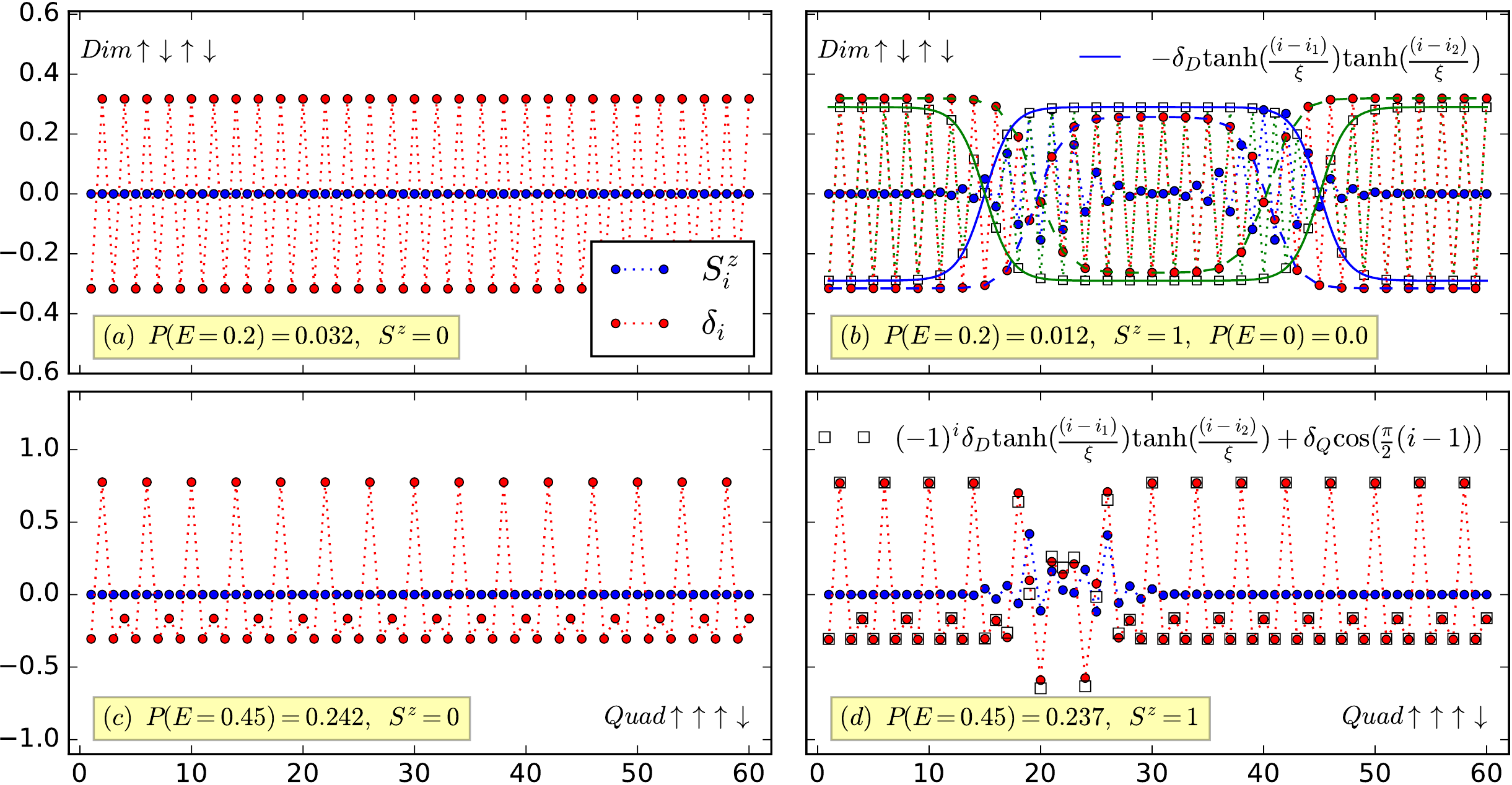}
   \caption{
   Distortion and magnetization profiles at the $M=0$ plateau for $E \neq 0$ 
   ($J_m=1$, $J_e=0.5$, $\alpha=1$ and $\beta=0.2$, in a $N=60$ sites chain with periodic boundary conditions).  The differences in local dipolar profiles are clearly seen for $E$ below or above $E_{c1}$ 
   (panels $(a)$ and $(c)$). 
   The soliton-antisoliton pair for the first magnetic excitation is glued together by the electric field (panels $(b)$ and $(d)$); blue and green curves in panel $(b)$ describe the repulsive soliton-antisoliton pair for $E=0$, added for comparison.
   }
\label{fig:4}
\end{figure}

The presence of a finite electric field $E<E_{c1}$  
penalizes the region with dipoles 
and distortions having the same sign (see Eq.\ (\ref{eq:H-fields})), 
gluing the soliton-antisoliton pairs and producing damping in the polarization switch-off effect (see Fig.\ \ref{fig:4}, 
upper right panel).  

\begin{figure}[ht!]
\centering
   \includegraphics[width=0.9\columnwidth]{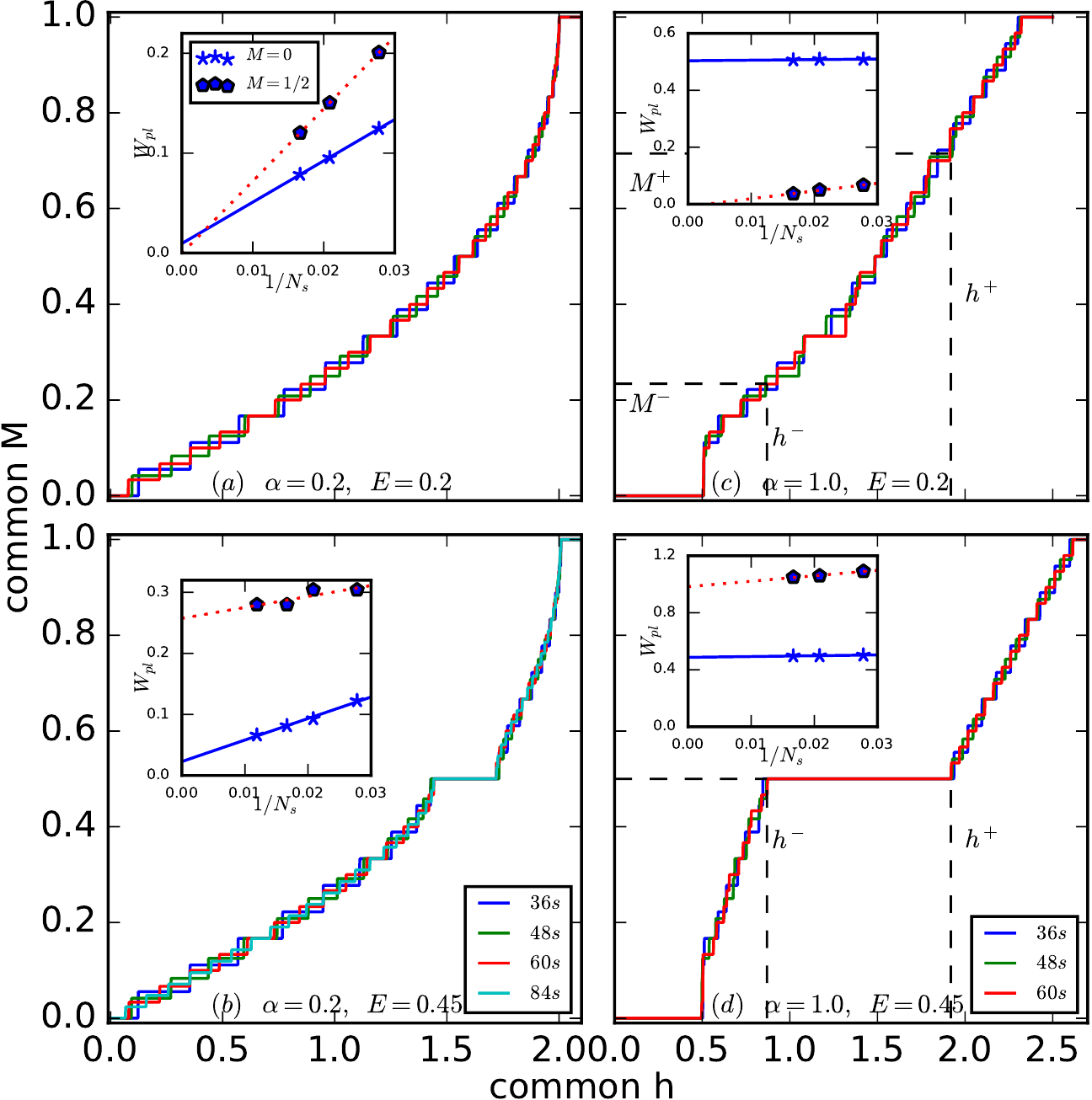}
   \caption{Magnetization curves for $E \neq 0$ (setting $J_m=1$, $J_e=0.5$ and $\beta=0.2$). A plateau at $M=0$ is always present; when $E$ drives the dipolar system into a quadrumerized phase a second plateau opens at $M=1/2$.
   $h^\pm$, the lower and upper boundaries of the $M=1/2$-plateau, are marked for later discussion 
   (see Fig.\ \ref{fig:6}). 
   }
\label{fig:3}
\end{figure}

Higher electric fields $E_{c1}<E<E_{c2}$ induce dipole flips, 
driving the electric subsystem to a $\uparrow \uparrow \uparrow \downarrow$ configuration. 
Being the distortions a superposition of 
period two  and  four  harmonics, 
the presence of magnetization plateaux at $M=0$ and $M=1/2$ is anticipated. 
We have checked numerically that this picture remains qualitatively the same 
when the dipolar subsystem is coupled with magnetism ($\alpha \neq 0$), 
with a smooth renormalization of the phase boundaries in Fig.\ \ref{fig:1}(d).
Representative magnetization curves exhibiting plateaux, computed numerically from DMRG, are shown in Fig.\ \ref{fig:3},
for  values of $E=0.2,\ 0.45$ and $\alpha=0.2,\ 1.0$.
One observes that the plateau at $M=0$ is always present, while a second plateau opens at $M=1/2$ when $E$ drives the dipolar system into the quadrumerized phase.
The plateaux widths are enhanced by higher magnetoelastic coupling $\alpha$.

Details on the quantum states at the $M=0$ plateau and their magnetic excitations
are shown in Fig.\ \ref{fig:4}.
We show  distortion and magnetization profiles for low electric fields, at $S^z=0$ (a) and first magnetized excitation (b). 
In the latter the continuous lines indicate the soliton profiles for $E=0$,
to be compared with the finite field profiles (dashed lines) that show the gluing of solitons. 
This gluing effect is more pronounced for electric fields in the quadrumerizad phase (c), as seen in panel (d) where distortions are fitted with 
%
$\delta_i = (-1)^i \delta_D \tanh(\frac{i-i_1}{\xi})\tanh(\frac{i-i_2}{\xi}) + \delta_Q\cos(\frac{\pi}{2}(i-1))$, with $i_{1,2}$ indicating the soliton positions.

%
\begin{figure}[ht!]
\centering
   \includegraphics[width=1.\columnwidth]{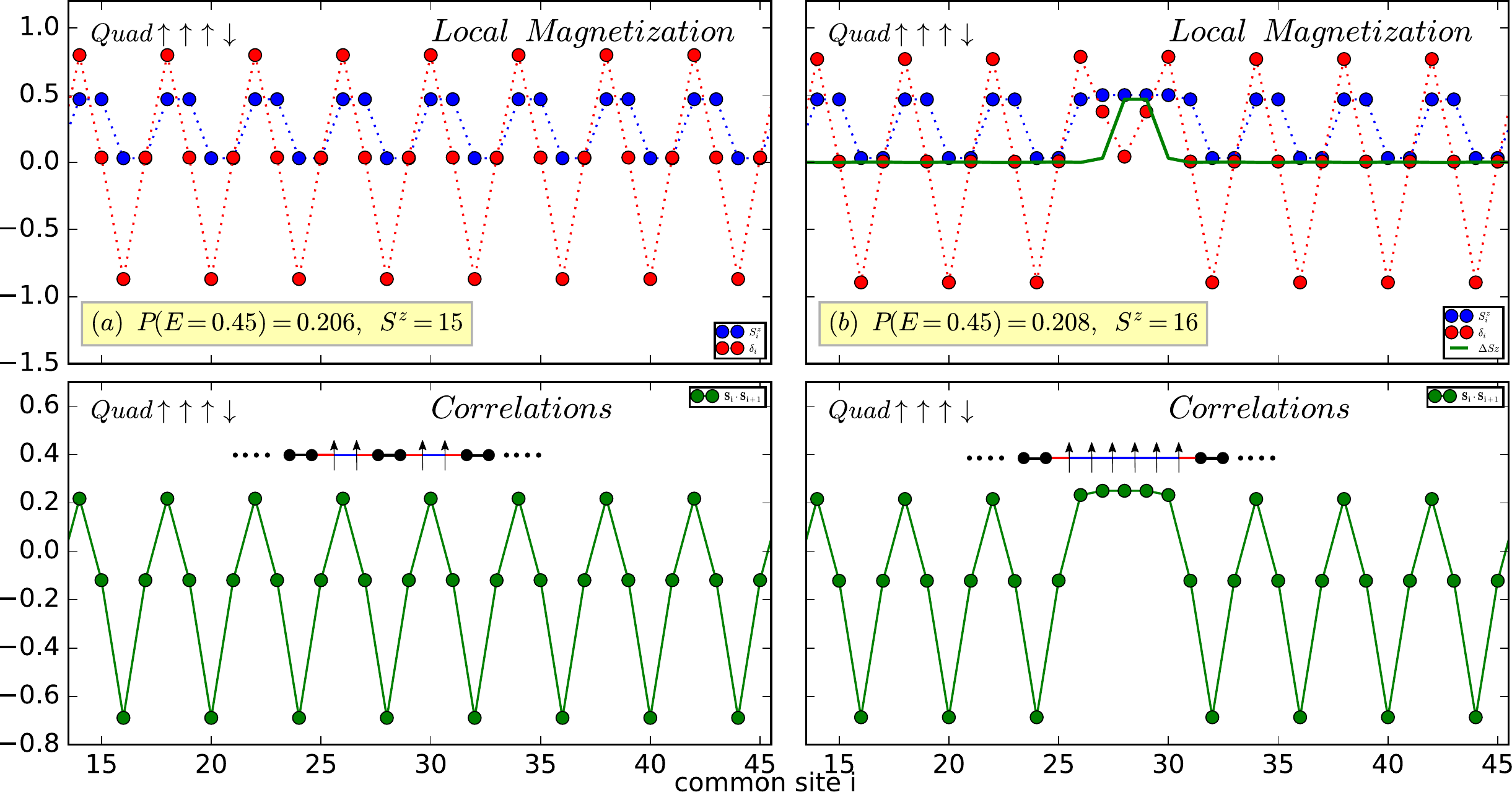}
   \caption{Quantum state signatures of the $M=1/2$ plateau (parameters as in Fig.\ \ref{fig:4}, $E=0.45$; we show again results for 60 sites, zoomed to visualize clearly the ordered direct product of singlets and spin up sites). 
   Local magnetization and nearest neighbor correlations are compatible with a 
   factorized quantum state of alternating up-up and singlet states. 
This is depicted graphically by black, red and blue bonds indicating respectively singlet, weak antiferro and ferromagnetic correlations.
   The lowest energy spin excitation is a localized triplet; the green line in panel (b) indicates the local increase in $S^z$.  
}
\label{fig:5}
\end{figure}

The plateau at $M=1/2$ has 
particular features not present in the spin-Peierls  one
at $M=0$. 
On the one hand, the magnetic wave function is compatible with an ordered direct product of singlets and spin up sites, 
as depicted in Fig.\ \ref{fig:5}. Magnetic excitations are simply given by magnons, that is singlet-triplet transitions 
that do not decay into solitons (see Fig.\ \ref{fig:5}, right panels). On the other hand the quantum state is 
topologically non trivial, as signaled by the even degeneracy of the entanglement spectrum \cite{Oshikawa2010}.

\begin{figure}[ht!]
\centering
   \includegraphics[width=1.0\columnwidth]{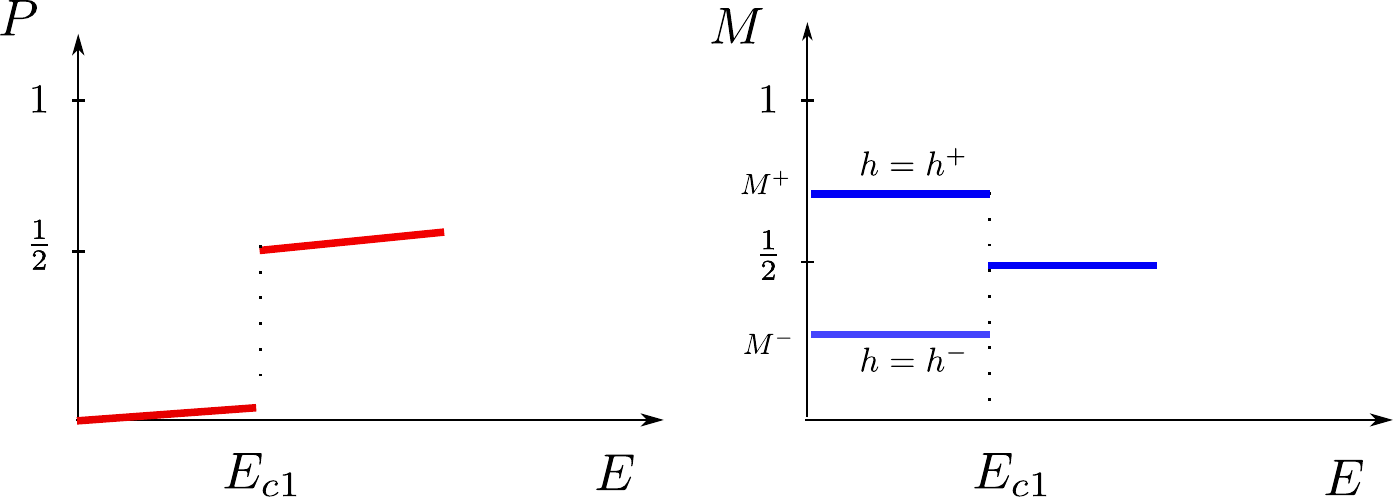} 
   \caption{Magnetoelectric response to the electric field in the quadrumerized scenario (schematic). 
   Under appropriate applied magnetic fields $h^\pm$, as the electric field produces a polarization jump (see Fig.\ 
\ref{fig:1}(d) at $J_e =0.5$) the magnetization switches from $M^\pm$ to $M=1/2$, the value at the plateau (see 
Fig.\ \ref{fig:3}).
   }
\label{fig:6}
\end{figure}

The present pantograph model also describes the effects of an electric field on the system magnetization. 
Let us analyze the scenario in which both dimerized and quadrumerized phases appear as a function of $E$, 
{\it e.g.} by choosing $J_e =0.5$, $\beta = 0.2$ (see Fig. \ref{fig:1}(d)). 
For $E_{c1} < E < E_{c2}$ the dipolar sector is quadrumerized and so is the lattice, 
which forces the magnetic sector to open a plateau at  $M=1/2$, as clearly seen 
from the numerical results in Fig.\ \ref{fig:3}. 
Choosing a background magnetic field $h^-$ at the lower boundary of this plateau, 
the magnetization will jump from some value $M^-<1/2$ to $M=1/2$ as the electric field crosses $E_{c1}$ from below; 
conversely, choosing  $h^+$ at the upper boundary 
the magnetization will jump from some value $M^+>1/2$ to $M=1/2$. This ME response is schematically depicted in Fig.\ \ref{fig:6}.
Such control of magnetization by an electric field is one of the goals of multiferroic technology developments 
\cite{techno}. 

Several quasi-one-dimensional materials showing multiferroicity have been studied in the last years 
\cite{Giovannetti,Streltsov,Khomskii2,ANNNI-models}. 
In most of these systems, 
a similar mechanism to the one proposed here seems to be relevant to describe the origin of the magnetoelectricity, 
though the spin ordering in some of them is of the type $\uparrow\uparrow\downarrow\downarrow$ 
at zero magnetic field, 
spins may have a strong Ising anisotropy or take alternating different values  along the relevant chain directions, etc. 
In order to describe these observed phenomena, 
one needs to consider further neighbors couplings between the neighboring spins and allow for higher 
spin  $SU(2)$ representations or even consider Ising spins.

In the cases in which the magnetic moments can be treated as Ising variables, 
such as Ca$_3$CoMnO$_6$, the ANNNI model has been proposed to describe the physics \cite{ANNNI-models}. 
Even in such cases, 
the description of the process of magnetic depolarization must include excitations and/or quantum fluctuations. 
In this respect, our model is expected to provide the correct description of the transition 
and could be tested against experiments. 

The extended $J_1-J_2$ model studied in \cite{Totsuka} shows an $M=1/2$ plateau with period four symmetry breaking
and 
dissociation of solitons as one increases or decreases the magnetization out of the plateau. 
In experiments done in R$_2$V$_2$O$_7$ (R=Ni,Co) a similar situation has been observed, together with a sharp change in $P$ 
on both sides of the 1/2 plateau \cite{RVO}. 
In spite of these differences, the behavior of the magnetization and electric polarization in a 
magnetic field for spin gapped phases even at non-zero field (plateaux phases) 
seems to be ubiquitous in all of the materials listed above.

The present mechanism is readily generalized to higher dimensions by considering 
the relevant structural units as octahedra in perovskites, double tetrahedra in 
hexagonal manganites, etc. 
These units containing the magnetic atoms are arranged, say, in the corners of a square/cubic lattice and 
a kind of spin-Peierls mechanism can occur. 
Linking again the deformation of the lattice along a given preferential direction 
with the height of the basic unit (see Fig.\ \ref{fig:1}(a)) 
the magnetoelectric coupling arises in the same way.    
Even in the case that tunneling between double-well potential minima was not negligible, 
and electric dipoles  were better described by a transverse Ising model, 
we expect our main conclusions to remain valid. Also higher spin magnetic ions, either classical or quantum, could be considered. 

Though the relation between striction and multiferroicity in quasi-one-dimensional systems has been discussed in several 
works \cite{Khomskii, Dagotto3,Dong-2009}, 
in most of the cases 
dipolar moments are not included as dynamical variables. 
In the present Letter, we fill this gap by proposing a more general mechanism 
that includes electro-elastic couplings via the distortion dependence of both 
local dipolar strengths and their interactions.
The full Hamiltonian couples spins and electric moments via lattice deformations through 
the proposed pantograph-like effect.

We hope that the present pantograph mechanism will shed light on the understanding of the 
microscopic origin of ME coupling in type II multiferroics.

\vspace{1cm}

\noindent {\em Acknowledgements}: D.C.C. acknowledges useful discussions with M. Jaime, M.L. Medarde, 
M. M\"{u}ller and J. White. 
This work was partially supported by CONICET (Grants No. PIP 2015-813 and No. PIP 2015-364), Argentina.

\end{document}